\documentclass[fleqn,10pt]{wlscirep}

\usepackage[utf8]{inputenc}
\usepackage[T1]{fontenc}
\usepackage{bm}
\usepackage{dsfont}
\usepackage{hyperref}

\title{Retrieving the structure of probabilistic sequences of auditory stimuli from EEG data}

\author[1]{Noslen Hern\'andez}
\author[1]{Aline Duarte}
\author[2]{Guilherme Ost}
\author[3]{Ricardo Fraiman}
\author[1]{Antonio Galves}
\author[4,*]{Claudia D. Vargas}
\affil[1]{Instituto de  Matem\'atica e Estat\'\i stica, Universidade de S\~ao Paulo, Brazil}
\affil[2]{Instituto de Matem\'atica, Universidade Federal do Rio de Janeiro, Brazil}
\affil[3]{Centro de Matem\'atica, Universidad de la Rep\'ublica, Uruguay}
\affil[4]{Instituto de Biof\'\i sica Carlos Chagas Filho, Universidade
  Federal do Rio de Janeiro, Brazil}
\affil[*]{cdvargas@biof.ufrj.br}
\date{November 28, 2019}
\providecommand{\mykeywords}[1]{\textbf{Keywords:} #1}

\begin{abstract}
Using a new probabilistic approach we model the relationship between sequences of auditory stimuli generated by stochastic chains and the electroencephalographic (EEG) data acquired while 19 participants were exposed to those stimuli. The structure of the chains
generating the stimuli are characterized by rooted and labeled trees whose leaves, henceforth called {\sl contexts}, represent the
sequences of past stimuli governing the choice of the next stimulus. A classical conjecture claims that the brain assigns probabilistic models to samples of stimuli. If this is true, then the context tree generating the sequence of stimuli should be encoded in the brain activity. Using an innovative statistical procedure we show that this context tree can effectively be extracted from the EEG data, thus giving support to the classical conjecture.
\end{abstract}

\begin{document}
	
	\flushbottom
	\maketitle
	%
	%
	\thispagestyle{empty}
	
\mykeywords{structural learning, context tree model, auditory evoked potential}
	
\section*{Introduction}

It has long been conjectured that the brain learns statistical regularities from sequences of stimuli \cite{VonHelmholtz:67}. The ability to extract and represent these regularities over time, often termed as statistical learning \cite{SCHAPIRO2015501,CONWAY2020279} plays a crucial role in perception and decision-making \cite{summerfield_expectation_2014,Armstrong_2017,de_lange_how_2018}.

Electrophysiological evidence of statistical learning has been often assessed by identifying signal differences for standard and unexpected stimuli embedded in a sequence \cite{dehaene_neural_2015}, or by looking for fluctuations in the measured signal associated with different surprise levels \cite{bornstein_dissociating_2012, maheu_brain_2019,lieder_neurocomputational_2013}. Much of this research has focused on the identification in the brain activity of signatures of either item frequency or transition probabilities associated to the sequence of stimuli \cite{grill-spector_repetition_2006,garrido_mismatch_2009,summerfield_neural_2008,todorovic_prior_2011,todorovic_repetition_2012,meyniel_human_2016,mittag_transitional_2016,naatanen_mismatch_2007}.

Here we address the conjecture that the brain learns statistical regularities from sequences of stimuli employing the new probabilistic framework introduced in Duarte et al. (2019) \cite{duarte_retrieving_2019}. In that framework, participants are exposed to a sequence of auditory stimuli while their electroencephalographic (EEG) activity is measured. The EEG signal is then segmented in the following way. The first EEG segment is the one recorded during the presentation of the first auditory stimulus; the second EEG segment is the one recorded during the presentation of the second auditory stimulus, and so on. Doing so, we now have two coupled sequences: the original sequence of auditory stimuli and the derived sequence of EEG segments recorded during the presentation of the sequence of the successive auditory stimuli. The question is whether the sequence of EEG segments expresses in some way the type of dependence from the past characterizing the sequence of auditory stimuli. A crucial point here is that the length of the relevant sequence of past choices of auditory stimuli governing the transition to the next one does not need to be fixed: it can very well depend on the sequence of past symbols itself. This framework allows to address von Helmholtz's classical conjecture as follows. Can the information concerning the sequences of past auditory units governing the successive transitions be retrieved from the statistical features of the EEG segments?

This approach is based on a genial intuition presented in Rissanen
(1983) \cite{rissanen_universal_1983}. He noted that in most sequences
of data observed in the real world, each new data unit is chosen in a
probabilistic way. This choice is made by taking into account a sequence
of past units whose length is not fixed, but changes as a function of
the sequence of past units itself. Following Rissanen
\cite{rissanen_universal_1983} we call a \textit{context} the sequence
of past units containing the relevant information required to generate
the next unit. The set of contexts defines a partition of the set of
all possible past sequences and can be represented by a rooted and
labeled oriented tree, henceforth called \textit{context tree}. To
completely characterize the way the sequence of units is generated, a
transition probability governing the choice of the next unit is
associated with each context.

Stochastic sequences as described above are called context tree models \cite{rissanen_universal_1983,Galves-Loch:08}. These models are able to approximate any stationary stochastic chain in an economic way, by using a small number of parameters and thus have been successfully employed to model biological and linguistic phenomena \cite{Buhlmann-Wyner:99,csiszar_context_2006,Leonardi06, GarivierLeonardi11,galves:12,Galves-Loch:13,BelloniOliveira17}.  

In Rissanen's words, von Helmholtz's classical conjecture can be rephrased as follows. Is it possible to retrieve the context tree generating the sequence of auditory stimuli from the sequence of EEG segments? With this formulation a major challenge is how to identify the statistical features of the EEG segments associated with each context. Duarte et al. (2019) \cite{duarte_retrieving_2019} addresses this question from an innovative mathematical point of view. By employing the statistical procedure proposed in Duarte et al. (2019)\cite{duarte_retrieving_2019}, we show that context trees generating sequence of auditory stimuli can effectively be extracted from EEG data.

\section*{Methods}
	
\subsection*{Participants}

Nineteen individuals (18 right-handed, 9 female, age average 30 and standard deviation 6.8 years) with no reported neurological, neuropsychiatric or auditory disease participated in the study. Participants consisted of 3 undergraduate and 4 graduate students as well as 13 health service professionals (physical therapists and psychologists), recruited from the staff of the Laboratory of Neuroscience and Rehabilitation of the Institute of Neurology Deolindo Couto. The menstrual cycle of the female participants was not considered. Participants signed an informed consent term. Research was performed in accordance with the relevant guidelines and regulations and approved by the Research Ethics committee of the Institute of Neurology Deolindo Couto at the Federal University of Rio de Janeiro (Plataforma Brasil process number 22047613.2.0000.5261).
	
\subsection*{Stimuli and task}

The auditory stimuli consisted of strong beats, weak beats, and silent
units presented at stimulus onset asynchrony of 450 ms. Strong and
weak beats consisted of hand claps (spectral frequency range: 0.2 to
15 KHz and maximal duration 200 ms each) recorded through a microphone.  
These signals were then digitized and sharp cut to a duration of 400 ms using the software Audacity
(Version 2.0.5.0). Presentation software (Sound card: SoundMAX HD Audio) was used to present the sequence of stimuli.

Participants sat in a comfortable chair in a dimly illuminated room. No task was asked for the participants to perform. They 
were informed neither about the intent of the study, nor about the existence of statistical regularities in the sequence of
stimuli. They were only instructed to close their eyes and listen attentively to the
sequences of auditory stimuli. Sounds were presented binaurally 
via headphones at around 60 dB SPL. The loudness of the
stimuli was individually regulated before the experiment start by
asking each participant to adjust the beat sounds
to a comfortable level.

The manner by which the sequences of stimuli were generated is
presented in the next sub-section (Structure of the sequences of
stimuli). Samples of the employed sequences of stimuli can be found in
the Suplementary Audio online.
	
\subsection*{Structure of the sequences of stimuli}

At this point, it is convenient to introduce a few formal definitions.
Let $A$ be a finite set. Given two integers $m,n$ with $m\leq n$, the
string $(u_m,\ldots, u_n)$ of symbols in $A$ will be denoted
$u^n_m$; its \textit{length} is $\ell(u^n_m)=n-m+1$. We will also use
the shorthand notation $u$ to denote the sequence $u^n_m$.

Given two sequences $u$ and $v$ of elements of $A$, $uv$ denotes the
sequence of length $\ell(u)+\ell(v)$ obtained by concatenating $u$ and
$v$. The sequence $u$ is said to be a \textit{suffix} of $v$ if there
exists a sequence $s$ satisfying $v=su$. When $v\neq u$ we say that
$u$ is a \emph{proper suffix} of $v$.

A finite set $\tau$ of sequences of elements of $A$ is a
\textit{context tree} if it satisfies the following conditions:
\begin{enumerate}
\item \textit{Suffix Property.} No $w\in \tau$ is a proper suffix of
  another element $u \in \tau$.
\item \textit{Irreducibility.} No sequence belonging to $\tau$ can be
  replaced by a proper suffix without violating the suffix property.
\end{enumerate}

The elements of $\tau$, i.e., the \textit{leaves} of the tree, are called \textit{contexts}. 

The \textit{height} of the tree $\tau$ is defined as $
\ell(\tau)=\max\{\ell(w) \,:\, w \in \tau\}.  $ For any sequence $u$
with $\ell(u)\geq \ell(\tau)$, we write $c_{\tau}(u)$ to denote the
only context in $\tau$ which is a suffix of $u$.

Let $\tau$ be a context tree and $p=\{p(\cdot\mid w): w\in\tau\}$ be a
family of probability measures on $A$ indexed by the contexts
belonging to $\tau$. The pair $(\tau,p)$ is called a
\textit{probabilistic context tree} on $A$.

A stochastic chain $(X_n: n \geq 0)$ taking values in $A$ is called a
\textit{context tree model} compatible with $(\tau,p)$ if
\begin{enumerate}
  \item $\tau$ defines a partition of the set of all possible
    sequences of past symbols generated by the chain;
\item \label{def.CTM} for any $n\geq \ell(\tau)$ and any finite
  sequence $x^{-1}_{-n}\in A^{n}$ such that
  $P\big(X^{n-1}_{0}=x^{-1}_{-n}\big)>0$, we have
\begin{equation}
\label{def:scmvl}
P\Big( X_n=a\mid X^{n-1}_{0}=x^{-1}_{-n}\Big) =p\big(a\mid c_{\tau}\big(x^{-1}_{-n}\big)\big) \ \  \mbox{for all} \ a\in A;
\end{equation}
\item no proper suffix of $c_{\tau}\big(x^{-1}_{-n}\big)$ satisfies condition 2.
\end{enumerate}

Given a probabilistic context tree $(\tau,p)$, the algorithm to
generate a context tree model can be summarized as follows:
\begin{itemize}
  \item Start the sample choosing, in an arbitrary way, any context
    with length equal to $\ell(\tau)$;
  \item at each subsequent step $n$, identify the context
    $c_{\tau}\big(X^{n-1}_{0}\big)$ corresponding to the past sequence
    $X_0,...,X_{n-1}$;
    \item choose the symbol $X_n$ using the transition probability
      $p\big(\cdot|c_{\tau}\big(X^{n-1}_{0}\big)\big)$.
\end{itemize}	 
       
For more details on context tree models we refer the reader to the founding papers by Rissanen (1983)\cite{rissanen_universal_1983} and
by B\"uhlmann and Wyner (1999)\cite{Buhlmann-Wyner:99} and to the survey by Galves and L\"{o}cherbach (2008)\cite{Galves-Loch:08}.

\begin{figure}[p] \centering
	\includegraphics[width=\linewidth]{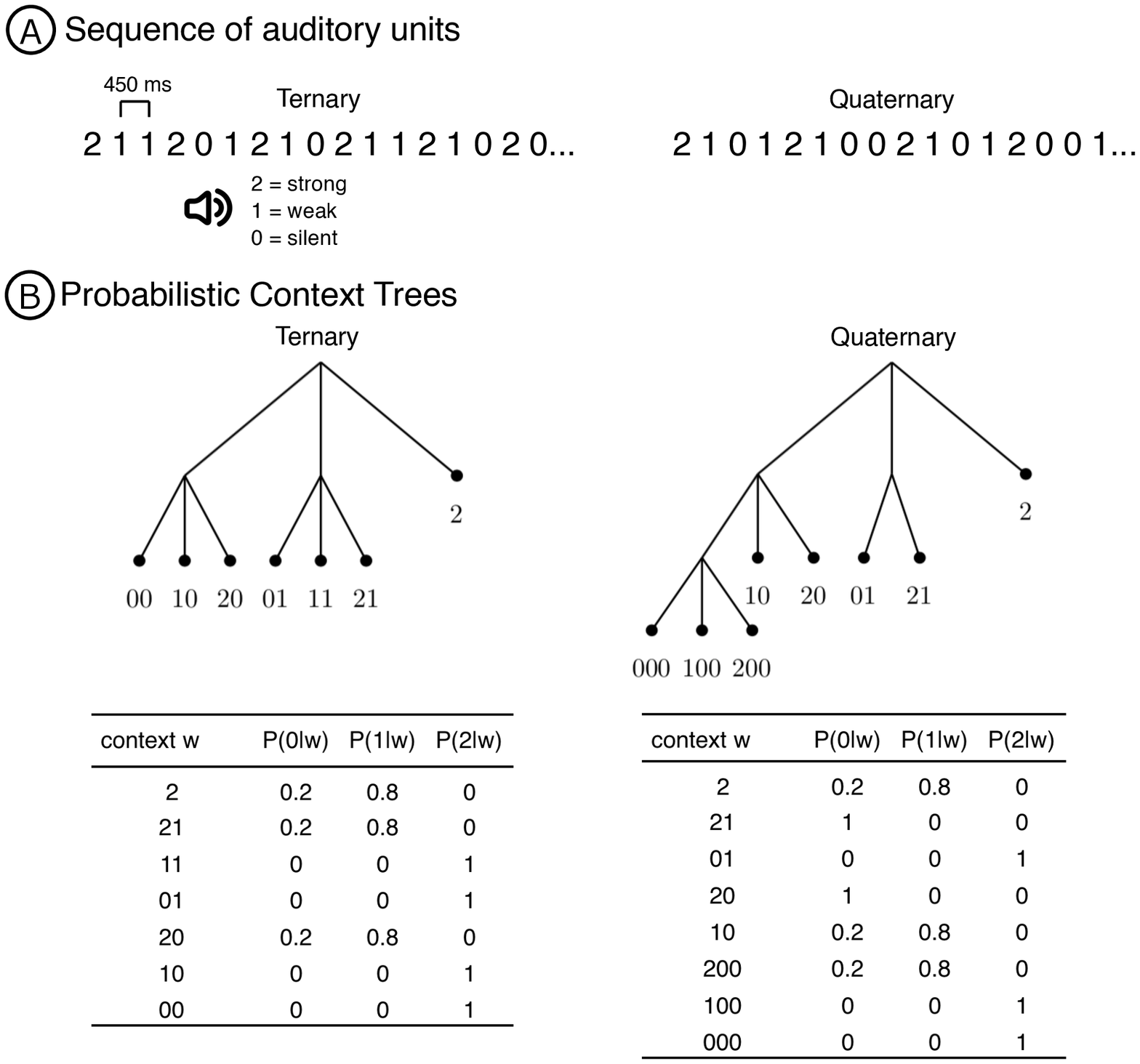}
	\caption{Experimental design. (A) Participants were
		exposed to sequences of auditory stimuli of three
		different types: strong beats, weak beats, and
		silent units, presented at every 450 ms. (B)
		Ternary and Quaternary context trees together with
		their associated family of transition probabilities.
		The transition probability associated to a context
		is used to choose the type of auditory stimulus
		appearing after the ocurrence of that context in the
		sequence of stimuli.}
        \label{fig:Design} 
\end{figure}

We can now describe in a precise way the context tree models used to generate the sequences of
stimuli. Take $A=\{0,1,2\}$, where 0 represents a silent unit, 1 represents a weak beat and 2 represents a strong beat (Figure
\ref{fig:Design}A). In our experiment, we used two different context tree models to generate the sequences of stimuli,
namely the \textit{Ternary} and \textit{Quaternary} conditions (Figure \ref{fig:Design}B). These two context trees reproduce mathematically two popular rhythms. The Quaternary condition is a simplified version of the Brazilian Samba. The Ternary condition is based on the Waltz structure.

In the Ternary condition, the tree of contexts is $\tau =
\{00,10,20,01,11,21,2\}$ and the family of associated probability
transitions is such that
 \begin{itemize}
 \item if the sequence of past symbols ends either with context 2, or
   with context 21, or with context 20, the next symbol will be either 0,
   with probability 0.2, or 1, with probability 0.8. The symbol 2 is
   not allowed to occur after contexts 2, 21 or 20;
 \item if the sequence of past symbols ends with context 11, or with
   context 10, or with context 01, or with context 00, the next symbol
   will be 2, with probability 1. The symbols 0 and 1 are not allowed
   to occur after contexts 11, 10, 01 and 00 (see the left lower table on
   Figure \ref{fig:Design}B).
   \end{itemize}
	 
In the Quaternary condition, the tree of contexts is $\tau =
\{000,100, 200, 10, 20,01,21,2\}$ and the family of associated
probability transitions is such that
 \begin{itemize}
 \item if the sequence of past symbols ends with context 2, or with
   context 10, or with context 200, the next symbol will be either 0, with
   probability 0.2, or 1 with probability 0.8. The symbol 2 is not
   allowed to occur after contexts 2, 10 or 200;
 \item if the sequence of past symbols ends with the context 21, or
   with the context 20, the next symbol will be 0, with probability
   1. The symbols 1 and 2 are not allowed to occur after contexts 21
   or 20;
  \item if the sequence of past symbols ends with context 01, or with
    context 100, or with context 000, the next symbol will be 2, with
    probability 1. The symbols 0 and 1 are not allowed to occur after
    contexts 01, 100 or 000 (see the right lower table in Figure
    \ref{fig:Design}B).
 \end{itemize}

Finally we consider sequences of stimuli generated by choosing at each step either 0 or 1 or 2 with equal probabilities 1/3, 1/3 and 1/3 independently of the past choices. This will be called the Independent condition. The goal of introducing this condition was to \textit{shuffle cards} before the participant is exposed to a next sample.  

For each condition, sequences of stimuli were generated independently
per participant. In this way, each participant was exposed to
different sequences of stimuli, all of them, however, being generated by the
same context tree model.
	 
\subsection*{Experimental design and procedures}

        Each participant was exposed to two 12.5 min blocks of
        sequences of auditory stimuli. The blocks were separated by a
        period of time ranging from 5 to 10 min, during which no data
        was acquired. Each 12.5 min block was composed by three 3.5
        min sub-blocks.  Each sub-block was a concatenation of three 1
        min sequences of auditory units generated using either the
        Ternary, the Independent or the Quaternary trees. Each
        sequence of auditory units was separated from the next one by
        a 15 seconds silent interval and each sub-block with three
        sequences was separated from the next one by a 1 min silent
        interval.
        
        For half of the volunteers the starting 12.5 min block
        contains sub-blocks of the type Ternary, Independent,
        Quaternary and the second 12.5 min block contains sub-blocks
        of the type Quaternary, Independent, Ternary. The inverse
        ordering was used with the other half, to balance possible
        order effects.
        	
\subsection*{Data preprocessing}
	
        EEG data was acquired at 250 Hz, using a 128 channels system
        (Geodesic HidroCel GSN 128 EGI, Electrical Geodesic Inc.).
        The electrode positioned on the vertex (Cz) was used as a
        reference during the acquisition. The electrode cap was
        immersed in saline solution (KCl) prior to data
        collection. During acquisition, the signal was amplified with
        a nominal gain of 20 times and analogically filtered
        (Butterworth first order band-pass filter of 0.1-200 Hz;
        Geodesic EEG System 300, Electrical Geodesic Inc.).
        
        The recorded signal was re-referenced to the average reference
        using NetStation software (Electrical Geodesic Inc) and
        filtered offline with a Butterworth fourth order band-pass
        filter of 1-30 Hz. EEG signal was then epoched with a
        peri-stimulus window of -50 to 400 ms (Figure \ref{fig:Analysis_Pipeline}A) using EEGLAB
        \cite{Delorme2004} running in MATLAB environment (Math-Works,
        Natick, MA, version R2012a). Baseline correction was performed
        employing the EEG signal collected from -50 to 0 ms before the
        auditory stimuli onset. The window of interest (from -50 ms to
        400 ms relative to the onset of the auditory stimulus) was
        chosen so as to include the duration of the auditory stimulus
        processing, spanning up to around 400 ms (see, for instance, Naatanen et al. 2001 \cite{Naatanen2001}; Garrido et al. 2009 \cite{garrido_mismatch_2009}).
        
\subsection*{Statistical data analysis}
  	
\begin{figure}[p]
	\centering
	\includegraphics[width=\linewidth]{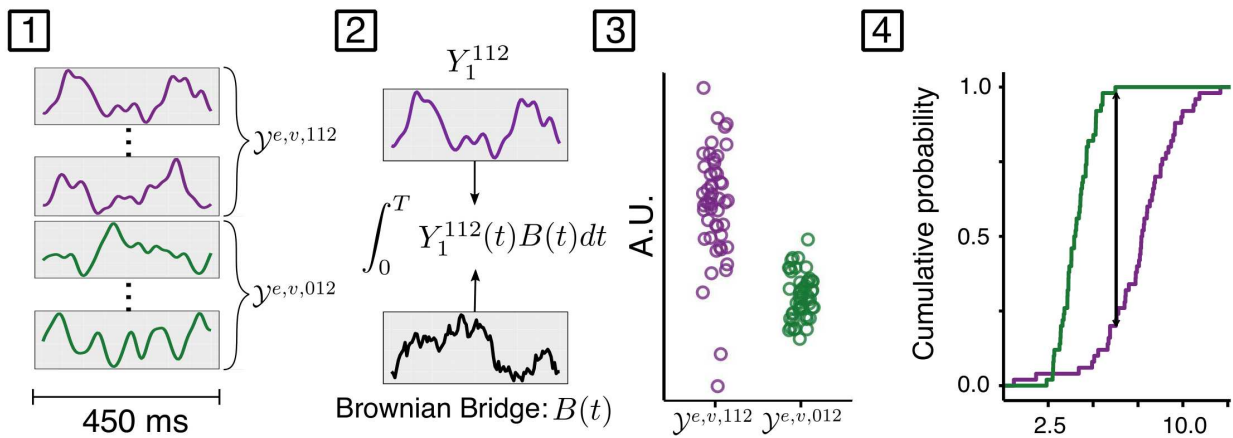}
	\caption{Context tree estimation pipeline. (A) EEG
		signal of one electrode segmented in windows of 450
		ms, starting 50 ms before the presentation of each
		stimulus (B) The pruning procedure is performed by
		(1) selecting not visited terminal subtrees of the
		complete tree, (2) testing the null hypothesis of
		equality of distributions for the sample of EEG segments
		associated to the leaves of that subtree, and (3)
		pruning the subtree if the null hypothesis is not
		rejected for all pairs of leaves in the subtree or
		keeping the subtree if the null hypothesis is
		rejected for at least one pair of leaves. (C) The
		test of the null hypothesis for the equality of distributions
		is performed by (1) selecting EEG segments
		associated with the pair of strings corresponding to
		the leaves that will be tested, (2) performing the
		inner product of the EEG segments with a realization
		of a Brownian bridge to obtain the projections, (3)
		having a distribution of the projections, (4)
		testing the null hypothesis for the equality of distributions
		for projections using the Kolmogorov-Smirnov test.}
	\label{fig:Analysis_Pipeline}
\end{figure}

The statistical procedure employed to estimate a context tree for each
participant and for each electrode was introduced in Duarte et
al. (2019)\cite{duarte_retrieving_2019}, to which we refer the reader
for a complete description of the method including the proofs of the
consistency result. This procedure is depicted in Figure
\ref{fig:Analysis_Pipeline}. 

Let $V$ be the set of all participants and $E$ the set of all
electrodes considered in the analysis. In our experiment the sets $V$
has 19 elements and $E$ is a set of 18 electrodes of the international
10-20 system
\[
E =\{\mbox{FP1}, \mbox{FP2}, \mbox{F7}, \mbox{F3}, \mbox{Fz},
\mbox{F4}, \mbox{F8}, \mbox{T7}, \mbox{C3}, \mbox{C4}, \mbox{T8},
\mbox{T7}, \mbox{P3}, \mbox{Pz}, \mbox{P4}, \mbox{P8}, \mbox{O1},
\mbox{O2}\}.
\]

For each participant $v \in V$ and each electrode $e \in E$, let
$X_1^{v},...,X_n^{v}$ be the sequence of stimuli the participant is
exposed to, and $Y_n^{e,v}$ be the segment of EEG data recorded during
the exposure to the auditory stimulus $X_n^v$. The sequence
$(X_1^v,Y_1^{e,v}),...,(X_n^v, Y_n^{e,v})$ will be used to estimate
the context tree $\hat{\tau}_n^{e,v}$. This is done by a pruning
procedure starting with an admissible context tree of maximal height
$k$ for the sequence of stimuli $X_1^v,..., X_n^v$, denoted by
$\mathcal{T}_n^k$, where $k$ is a suitable positive integer. 

Following Duarte et al. (2019) \cite{duarte_retrieving_2019}, an admissible context tree of maximal height $k$ is any irreducible tree of height $k$ in which all the contexts are finite sequences of symbols in the set $\{0,1,2\}$ that effectively occur in the sequence of stimuli. As an illustration, Figure \ref{fig:Analysis_Pipeline}B, step 1, shows an admissible context tree of maximal height 3 for the Ternary condition.

Before summarizing the pruning procedure, it is convenient to
introduce a new definition. A \textit{subtree} in a tree $\tau$ induced
by the sequence $u$ is defined as the set $B_{\tau}(u) = \{w \in \tau:
u \mbox{ is a suffix of } w\}$. The subtree $B_{\tau}(u)$ is called
\textit{terminal} if for all $w \in B_{\tau}(u)$ it holds that
$w = au$ for some $a \in A$ (see Figure \ref{fig:Analysis_Pipeline}B,
step 1).

Given a tree $\tau$ and $w = w_{-k}^{-1}\in \tau $, let $\mathcal{Y}^{e,v,w}$
be the set of EEG segments recorded during the occurrence of the auditory
unit $w_{-1}$, when it appears after the units $w_{-k}^{-2}$ (in
Figure \ref{fig:Analysis_Pipeline}A, EEG segments associated to the
leaves 012 and 112 are marked in green and purple, respectively).

The pruning procedure can be informally described as follows. 
\begin{enumerate}

\item Compute $\mathcal{T}_n^k$ which is the admissible context tree of maximal height $k$, for the
  sequence of stimuli $X_1^v,..., X_n^v$.
  
\item We start the pruning procedure from the admissible context tree $\tau = \mathcal{T}^k_n$.   

\item For each terminal subtree $B_{\tau}(u)$ on $\tau$ and for all
  pairs of leaves $au, bu \in B_{\tau}(u)$ with $a,b \in A$, test the
  null hypothesis that the distribution of the EEG chunks
  $\mathcal{Y}^{e,v,au}$ and $\mathcal{Y}^{e,v,bu}$ associated to the
  leaves $au, bu$ is equal (Figure \ref{fig:Analysis_Pipeline}B, step
  2).

\begin{enumerate}
	\item If the null hypothesis is not rejected for any pair of
          leaves $au, bu \in B_{\tau}(u)$, we conclude that the
          occurrence of the symbol $a$ or $b$ before the sequence $u$
          do not affect the distribution of EEG chunks and we prune
          the terminal subtree $B_{\tau}(u)$. Pruning the subtree
          $B_{\tau}(u)$ means updating the set $\tau$ by eliminating
          from it the elements in $B_{\tau}(u)$ and adding to it the
          sequence $u$ (Figure \ref{fig:Analysis_Pipeline}B, step
          3). Formally, $\tau = \tau \setminus B_{\tau}(u) \cup
          \{u\}.$
	
	\item If the null hypothesis is rejected for at least one pair
          of leaves $au$ and $bu$, we conclude that the distribution
          of EEG chunks depends on the entire sequence $au$ and $bu$
          and we keep the terminal subtree $B_{\tau}(u)$ in $\tau$
          (Figure \ref{fig:Analysis_Pipeline}B, step 3).
\end{enumerate}

\item Repeat step 2 until all terminal subtrees have been tested.

\item Call $\hat{\tau}_n^{e,v}$ the tree constituted by the sequences
  in $\tau$ when the pruning procedure ends.
\end{enumerate}		

Figure \ref{fig:Analysis_Pipeline}B exemplifies the pruning of the terminal subtree $\{012, 112\}$. At this step, we want to test if the distribution of the EEG responses to the string 2 preceded by the strings 01 and the EEG responses to the string 2 preceded by the string 11 are different. On one hand, if the distribution of the EEG responses are identical this means the last symbol in the past (i.e., 0 in the string 012 and 1 in the string 112) does not influence the EEG responses. In this case we discard the strings 012 and 112 in the tree, and replace them by the string 12. On the other hand, if the distribution of the EEG responses are not identical this means the last symbol in the past (i.e., 0 in the string 012 and 1 in the string 112) does influence the EEG responses. In this case the strings 012 and 112 are not pruned, becoming contexts of the tree. Notice that with the analysis of this subtree, we can not conclude that the EEG responses to token 2 are not influenced by two symbols backwards in the past. To do so, we need also to analyze the terminal subtree $\{002, 102\}$.

To test the equality of distributions for two samples of EEG segments, Duarte
et al. (2019)\cite{duarte_retrieving_2019} use the projective method
proposed in Cuesta-Albertos et al. (2006) \cite{Cuesta2006}. This method takes a function and
transforms it into a real number (i.e., its projection in a Brownian
Bridge). The theorem presented in Cuesta-Albertos et al. (2006) \cite{Cuesta2006} guarantees that if we
reject the null hypothesis that the distribution of the projections
corresponding to two samples of functions are equal, we can also
reject the null hypothesis that the two samples of functions have the
same distribution in the functional space. To obtain the projections
corresponding to the EEG segments, we perform an inner product of the
EEG function with a realization of a Brownian Bridge. The equality of
distribution for the resulting samples of real numbers is tested using the
Kolmogorov-Smirnov test, with alpha at 0.05 (Figure
\ref{fig:Analysis_Pipeline}C). We use the same realization of the
Brownian Bridge to project all EEG segments. To guarantee the
stability of the estimated context trees, we perform the projection
5000 times using different Brownians, which implies performing 5000
hypothesis tests. We only reject the null hypothesis that the distributions are
the same if more than 276 null hypothesis are rejected. This threshold
was derived according to a binomial distribution where the probability
of success corresponds to the probability of rejecting an individual
test assuming the null hypothesis is true, which is 0.05.

After estimating one context tree per electrode per
participant, $\hat{\tau}_n^{e,v}$, we ended up on each electrode
$e$ with a set of context trees $T_n^e= \{\hat{\tau}_n^{e,v}, v\in V\}$
corresponding to the context trees estimated for
all participants $V$ in that electrode. We summarized this set
of trees in a unique tree $\tilde{\tau}^e$ called \textit{mode context tree},
defined as follows:

The sequence $w$ is a context in $\tilde{\tau}_n^e$ if the
following conditions hold:
\begin{itemize}
	\item[i)] $\Sigma_{v \in V} \mathds{1}_{\left\{w \in \hat{\tau}_n^{e,v}\right\}} > \max_{u: w \mbox{ \tiny proper suffix of } u} \Sigma_{v \in V} 
	\mathds{1}_{\left\{u \in \hat{\tau}_n^{e,v} \right\}}$
	\item[ii)] condition (i) is not fulfilled for any sequence $u$ which is a proper suffix of $w$.
\end{itemize}

\begin{figure}[h]
	\centering
	\includegraphics[width = 0.9 \linewidth]{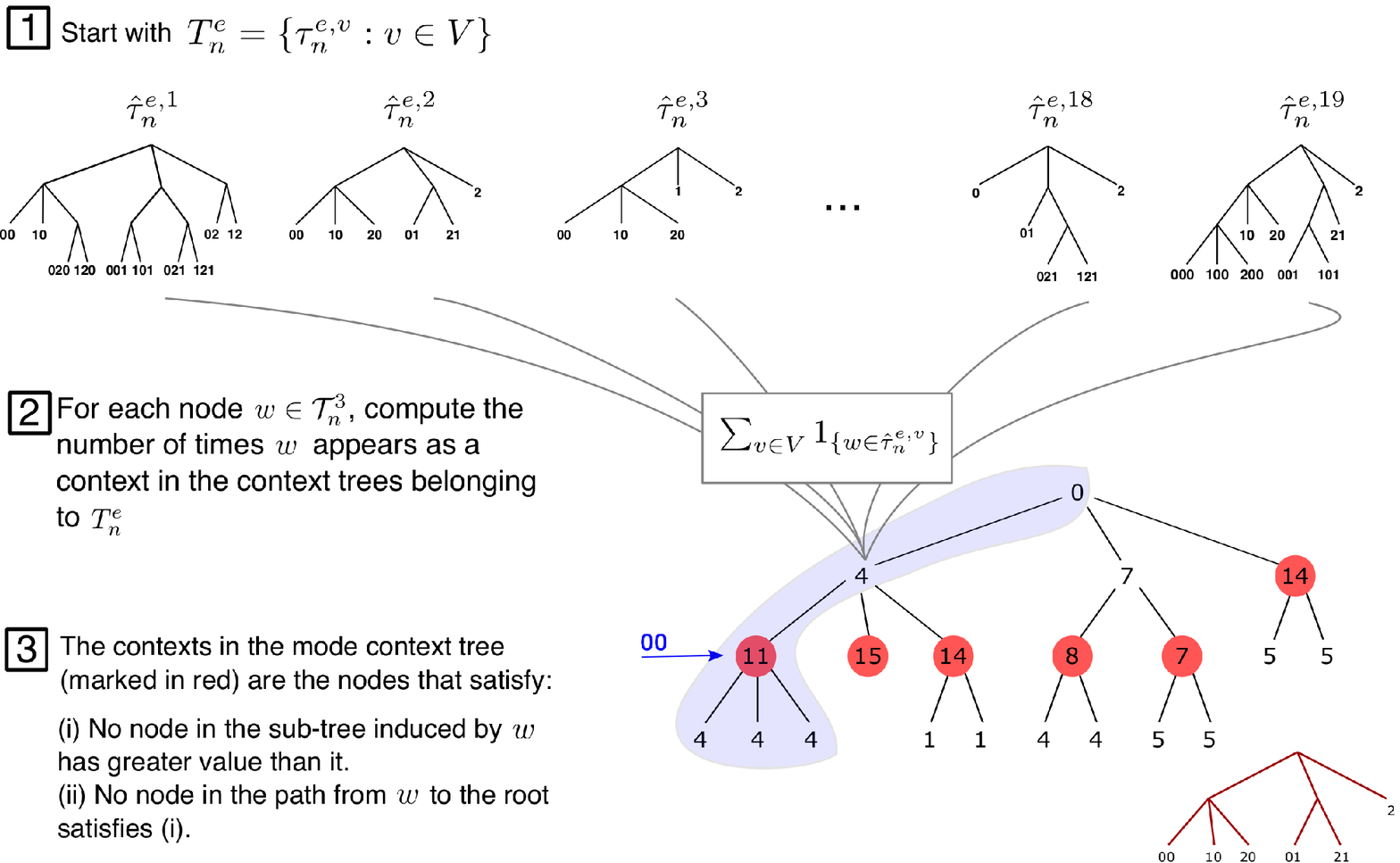}
	\caption{Mode context tree estimation. Start with the set $T_n^e$ which contains the context tree estimated from the data collected at electrode $e$ across participants. Then, for each node $w \in \mathcal{T}_n^3$, with $\mathcal{T}_n^3$ the admissible context tree of maximal height 3, we compute $\Sigma_{v \in V} \mathds{1}_{\{w \in \hat{\tau}_n^{e,v}\}}$ which is the number of times $w$ appears as a context in the context trees retrieved from data collected at electrode $e$ across participants. In the Figure these contexts are indicated in red. The contexts of the mode context tree are the nodes that satisfy i) no node in the subtree induced by $w$ has greater value than it; ii) no node in the path from $w$ to the root satisfies i). For instance, the node 00 appears 11 times, while the nodes 000, 100 and 200 which are in the subtree induced by 00 appear 4 times each. Besides, the ancestor nodes 0 and the root appear 4 times and zero time, respectively. Therefore, 00 appears as a context in the mode context tree.}
	\label{fig:mode_context_tree}
\end{figure}

In words, a sequence $w$ is a context in the mode context tree corresponding to an electrode $e$, when $w$ is more often a context in $T^e$ than any sequence for which $w$ is a suffix. Furthermore, no suffix of $w$ can satisfy the above condition. Thus, the leaves (i.e. contexts) of the mode context tree are those sequences that were identified as contexts more often across participants. Figure \ref{fig:mode_context_tree} exemplifies the computation of the mode context tree. 

\section*{Results}	

\subsection*{Retrieving context trees from EEG data}
			
We estimated a context tree per electrode from the EEG data collected while the participants were exposed to sequences of auditory stimuli generated by the Ternary and Quaternary conditions. For each condition and each electrode, the mode context tree summarizes the results obtained across participants.

\begin{figure}[p]
		\centering
		\includegraphics[width=0.9\linewidth]{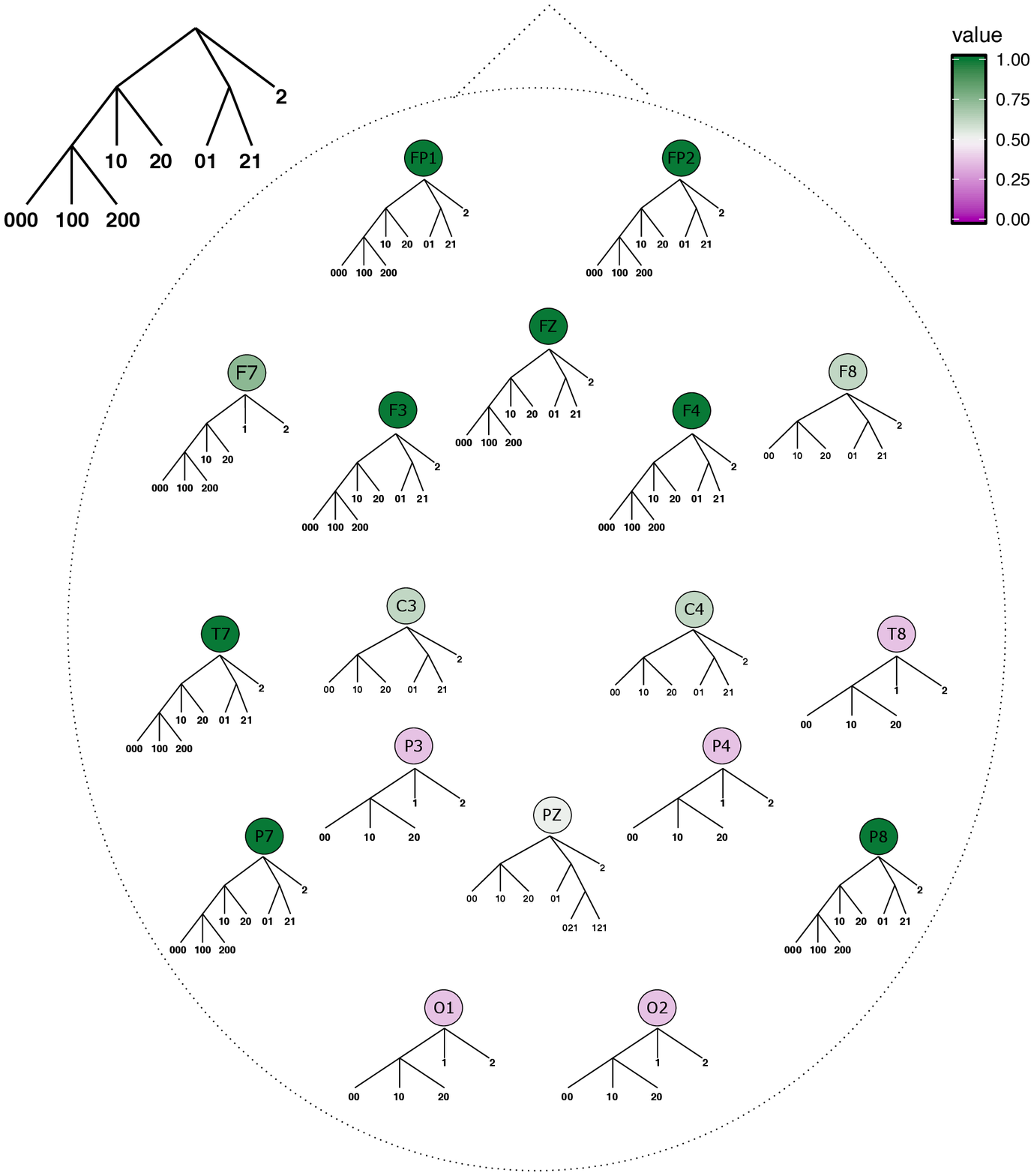}
		\caption{Mode context trees across electrodes for the
                  Quaternary condition. For each electrode, the mode
                  context tree represents the sequences that were more
                  often estimated as contexts across participants for
                  sequences of stimuli generated by the Quaternary
                  context tree. The color scale indicates the
                  similarity of a mode context tree $\hat{\tau}^e$
                  with the Quaternary context tree $\tau_{qua}$
                  computed as $s(\hat{\tau}^e, \tau_{qua}) = \Sigma_{w
                    \in \tau_{qua}} (1/8) \mathds{1}_{\{w \in
                    \hat{\tau}^e\}}$. The closer to one the similarity
                  value, the more similar the two trees are.}
		\label{fig:quaternary_tree}
	\end{figure}

The results obtained in the Quaternary condition are plotted in Figure \ref{fig:quaternary_tree}. In 5 frontal (FP1, FP2, FZ, F3, and F4), 2 parietal (P7 and P8) and one temporal (T7) electrode, the mode context tree exactly matches the tree that generated the sequence of stimuli, namely $\tau = \{000, 100, 200, 10, 20, 01, 21, 2\}$.

In addition, the estimated mode context tree only misidentified $1$ as a context in F7 and $00$ as a context in F8, C3 and C4. In PZ, the mode context tree misidentified $00$, $021$ and $121$ as contexts. Finally, for T8, P3, P4, O1 and O2, besides the lack of the third order contexts 000,100 and 200, the symbol 1 is misidentified as a context.

Summarizing, contexts 2, 10 and 20 were perfectly identified in all the 18 electrodes; contexts 01 and 21 were well identified in 11 out of 18 electrodes and the third order contexts 000,100 and 200 were identified in 9 out of 18 electrodes. 	
	
\begin{figure}[p]
		\centering
		\includegraphics[width=0.9\linewidth]{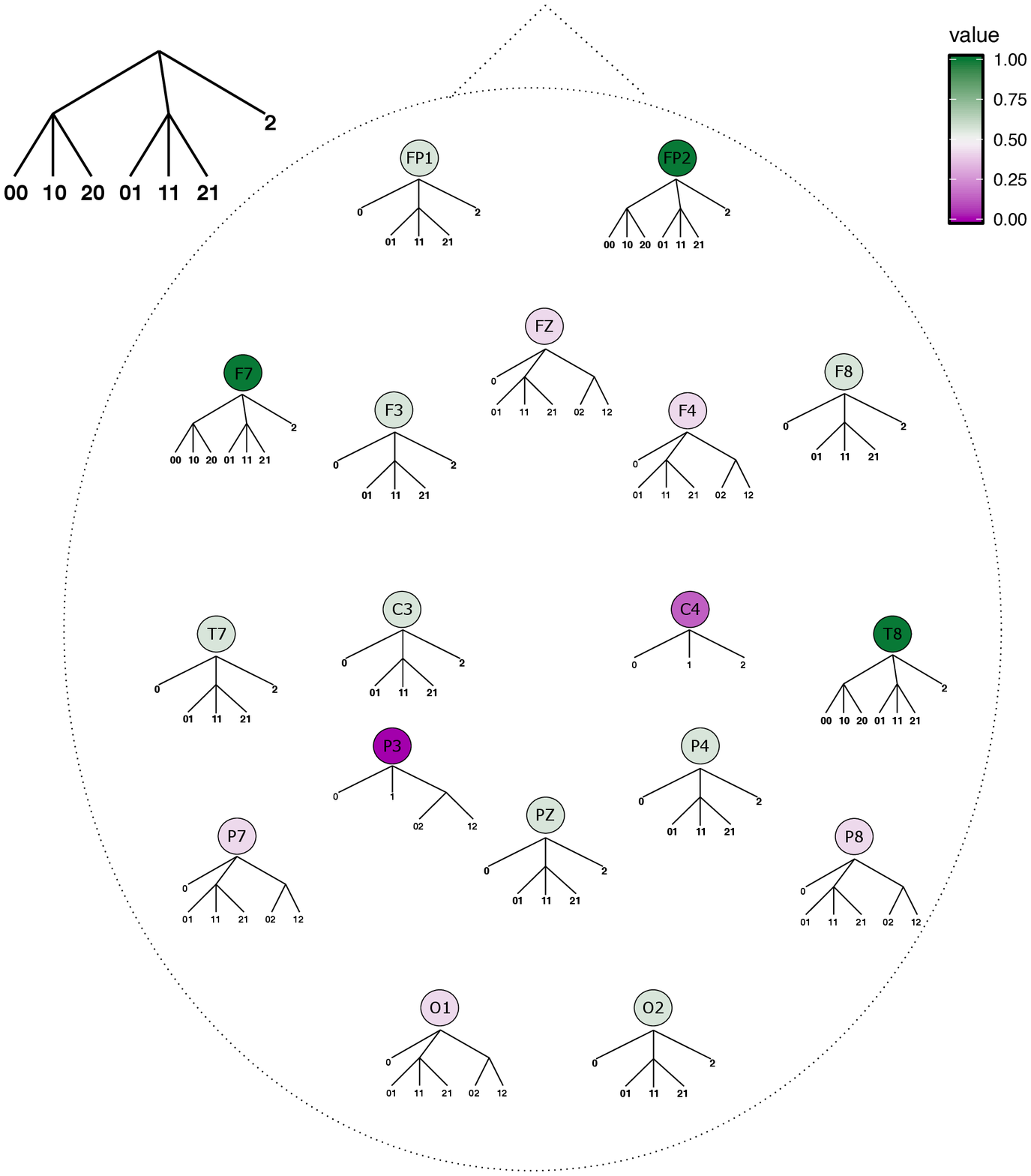}
		\caption{Mode context trees across electrodes for the
			Ternary condition. For each electrode, the mode
			context tree represents the sequences that were more
			often estimated as contexts across participants for sequences of
			stimuli generated by the Ternary context
			tree. The color scale indicates the similarity of
			a mode context tree $\hat{\tau}^e$ with the Ternary context
			tree $\tau_{ter}$ computed as $s(\hat{\tau}^e, \tau_{ter}) = \Sigma_{w \in \tau_{ter}} (1/7) \mathds{1}_{\{w \in
				\hat{\tau}^e\}}$. The closer to one the similarity value, the more similar the two trees are.}
		\label{fig:ternanry_tree}
	\end{figure}

Results for the Ternary condition are plotted in Figure \ref{fig:ternanry_tree}. For this condition, the mode context tree obtained for electrodes FP2, F7 and T8 was exactly the one that generated the stimuli, namely $\tau = \{00,10,20,01,11,21,2\}$. Moreover, 12 out of 18 electrodes (FP2, FP1, F7, F3, F8,T8, C4, C3, T7, PZ, P4, O2) had 2 as a context. Also, 16 out of 18 electrodes (FP1, FP2, F7, F3, FZ, F4, F8, T7, C3, T8, PZ, P4, P7, P8, 01, 02) contained the subtree $\{01,11,21 \}$. Surprisingly, the second order dependencies preceding the silent unit in the subtree $\{00,10,20\}$ were present only in 3 out of 18 mode context trees. Finally, inexistent second order dependencies appeared in electrodes FZ, F4, P7, P8 and O1.
	
\subsection*{Interindividual variability}

To evaluate the interindividual variability, for each electrode we
calculated the distance between the mode context tree and the context
trees retrieved across participants, as in Balding et al. (2009)
\cite{Balding09}. Informally speaking, the distance between two
context trees is the weighted average of the boolean variables
associated to each candidate node of the tree. To each node the
boolean variable assuming the value 1, if the node belongs to one of
the two context trees, but not to both, otherwise assuming the value
0. We refer the reader to Balding et al. (2009) \cite{Balding09} for a
formal definition.

\begin{figure}[h]
	\centering
	\includegraphics[width = \linewidth]{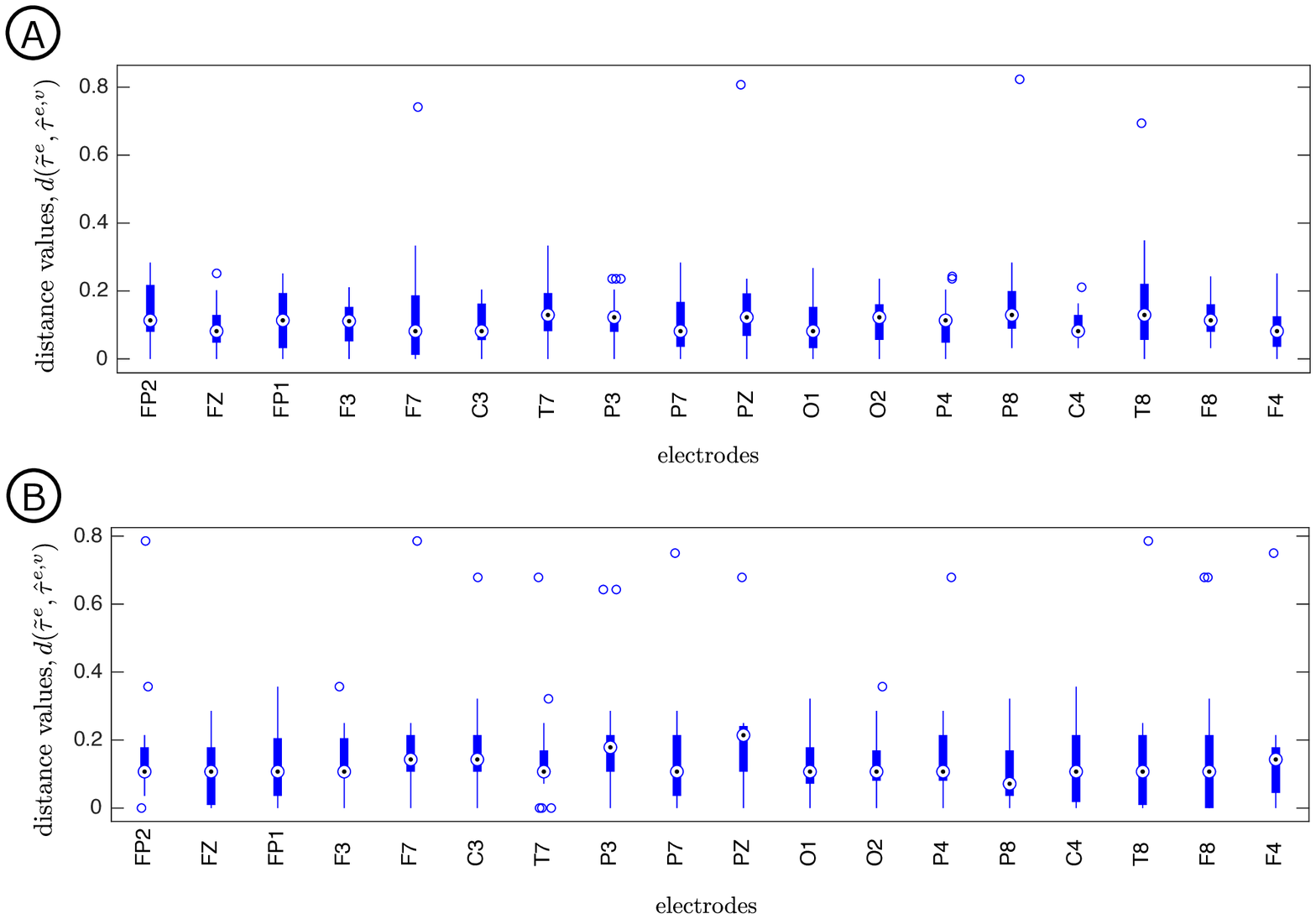}
	\caption{Boxplots depict the distance values between the mode context tree $\tilde{\tau}^e$ and the context tree retrieved per participant $\hat{\tau}^{v,e}, v \in V$ at a given electrode $e$ for the Quaternary (A) and Ternary (B) conditions.}
	\label{fig:variance_boxplot}
\end{figure}

Figure \ref{fig:variance_boxplot} depicts the boxplots of the set of distance values between the mode context tree and the context trees retrieved per participant at a given electrode for the Quaternary (A) and the Ternary (B) conditions.

It is evident from Figure \ref{fig:variance_boxplot} that the individual context trees retrieved per participant at a given electrode concentrate around the mode context tree. Moreover, both the median of the distances and the interquartile range are small. The fact that for each electrode the distances between the mode context tree and the retrieved context trees across participants are concentrated in a small interval around the median value indicates a relatively low interdindividual variability.

The complete sets of context trees estimated per participant and electrode for each condition are presented in Figures S5 to S40 in the supplementary material. Furthermore, the number of times each finite sequence of stimuli was estimated as a context across participants for each electrode on each condition are depicted in Figures S1, S2, S3 and S4 in the supplementary material.

\subsection*{Likelihood of obtaining the correct context tree by chance}

To evaluate how likely it is to obtain the results above just by chance we did a simulation experiment. This experiment assumes that the retrieved context trees have been chosen in a purely random and uniform way. In the Quaternary condition the pruning procedure could return a context tree among 51 possible candidates. The candidates are the context trees that could be produced by successively pruning subtrees in all possible cases, starting from the admissible context tree of maximal height 3. In the Ternary condition the set of possible candidates contains 46 context trees.  

For each condition, we simulate this situation by randomly choosing sets of 19 trees uniformly in the set of possible candidates context trees. We repeat this procedure 10.000 times. For each set of 19 randomly chosen context trees we compute the mode context tree.

\begin{figure}[p]
	\centering
	\includegraphics[width = 0.8\linewidth]{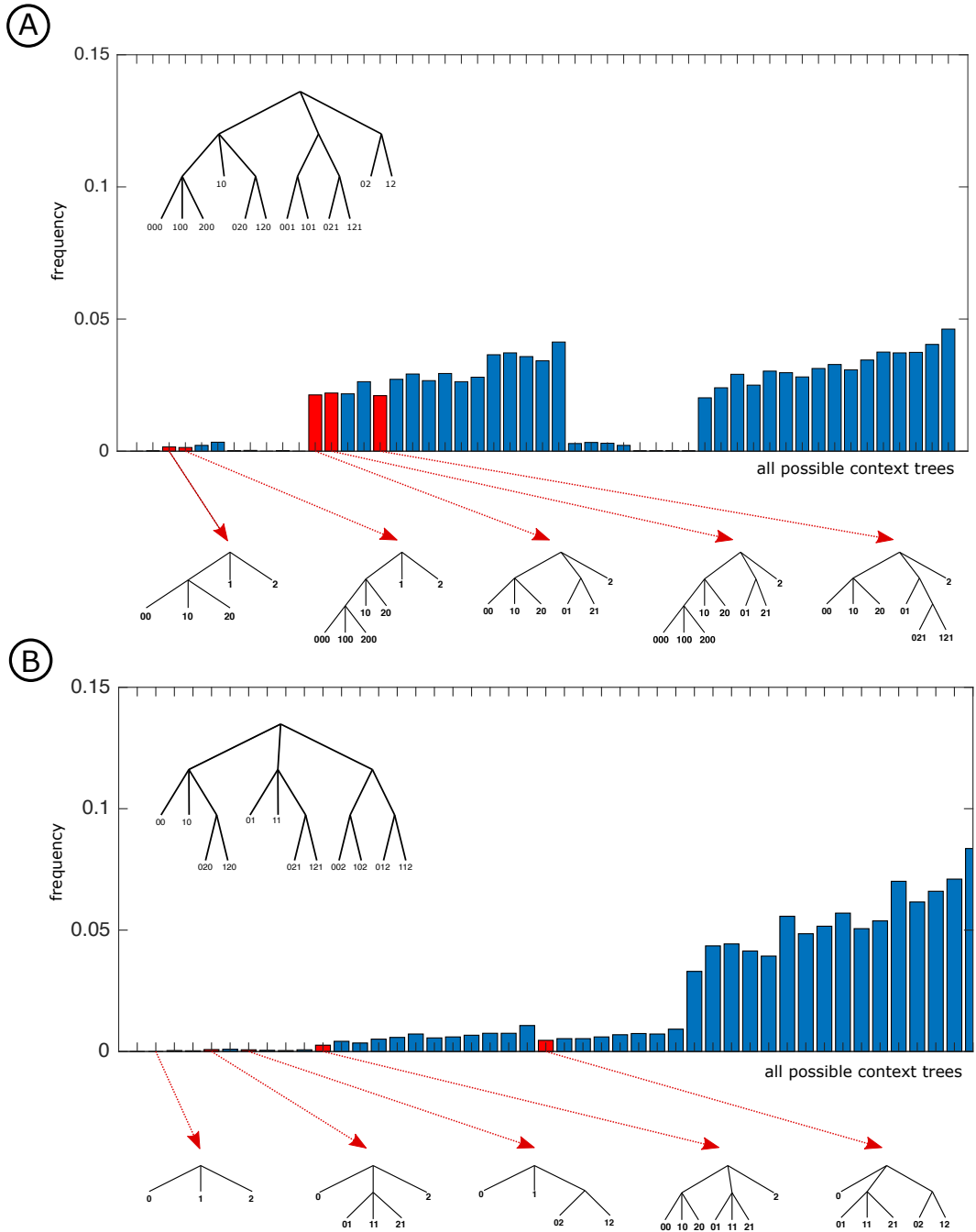}
	\caption{Empirical distribution of the mode context trees obtained by randomly choosing sets of 19 trees uniformly in the set of possible candidate context trees both for the Quaternary (A) and the Ternary (B) conditions. For each set of 19 randomly chosen context trees we compute the mode context tree. This procedure is repeated 10.000 times. In the histogram, the trees appearing from the experimental data analysis are indicated. The corresponding admissible context trees of maximal height 3 are depicted at the top of each plot.}
	\label{fig:prob_searchspace}
\end{figure}

Figure \ref{fig:prob_searchspace} presents the empirical distribution of the mode context trees obtained by means of the random choice described above for both conditions. In the Quaternary condition (Figure \ref{fig:prob_searchspace}A), the resulting empirical distribution assigns a probability close to 0.02 to the Quaternary tree that generated the sequence of stimuli.

If the result obtained with the EEG data collected in the Quaternary condition was produced by chance, as it was the case in the simulation, its likelihood could be easily computed as follows. As indicated in Figure \ref{fig:quaternary_tree}, the Quaternary tree appears as mode context tree across participants in 8 electrodes, namely FP1, FP2, FZ, F3, F4, T7, P7 and P8. Assuming that the data collected at each one of the 18 electrodes returns the Quaternary tree by chance and independently of the other electrodes with probability 0.02, then the probability that the Quaternary tree appears as mode context tree in exactly 8 out from 18 electrodes can be computed by a Binomial distribution with parameters 18 and 0.02. The probability that a Binomial distribution with parameters 18 and 0.02 takes the value 8 is equal to ${18 \choose 8} * (0.02)^8 * (0.98)^{10} \approx 1/10^{9}$, which is overwhelmingly small. Using the Binomial we can also compute the probability that the Quaternary tree appears at least 8 times among the 18 context tree modes. This probability is of the order of $1/10^{9}$. These results show that obtaining the Quaternary tree in at least 8 out of 18 context tree modes can hardly be explained by pure chance. Similar results can be obtained in the Ternary case.

\section*{Discussion}
      
In the present study, we employed a new statistical framework
\cite{duarte_retrieving_2019} to retrieve from EEG signals the
structure of the algorithm governing the production of a sequence of
auditory stimuli. For each electrode and condition the set of
context trees retrieved accross participants was summarized in the mode
context tree associated to each electrode and condition.

Both for the Quaternary and the Ternary conditions, the context trees
generating the sequences of stimuli were completely retrieved from the
EEG signals recorded in certain electrodes.  For other electrodes,
however, the retrieved mode context trees did not match exactly the
ones that generated the sequences of stimuli. Notwithstanding, some
contexts were estimated quite consistently both for the Quaternary and
the Ternary conditions. For instance, context 2 appeared in all mode
context trees for the Quaternary condition, and in 12 out of 18 mode
context trees for the Ternary condition.  In addition, the subtree
$\{01, 21\}$ appeared in the mode context trees obtained in 11 out of
18 electrodes for the Quaternary condition, and the subtree
$\{01, 11,21\}$ appeared in 16 out of 18 mode context trees for the
Ternary condition.  Finally, the subtree $\{000, 100, 200\}$ was
present in 9 out of 18 mode context trees in the Quaternary
condition. Likewise, the subtree $\{00,10,20\}$ was retrieved only in
3 out of 18 mode context trees in the Ternary condition.
	
Previous MEG/EEG and fMRI studies had shown that brain signals can be associated with transition probabilities of first order \cite{maheu_brain_2019,bornstein_dissociating_2012,kobor_erps_2018,Wang2017}. Furthermore, studies using serial reaction time tasks and artificial grammar learning had proven that humans can learn stimuli sequences that contain higher-order temporal dependencies \cite{cleeremans_implicit_1998,reed_assessing_1994,wang_learning_2017,cleeremans_learning_1991}. Our results show that it was possible to retrieve second ($\{01,21\}$) and third ($\{000, 100, 200\}$) order temporal dependencies from EEG signals in the Quaternary condition. Likewise, second order dependencies ($\{01, 11, 21\}$) could be retrieved in the Ternary condition. Furthermore, our approach allowed to go beyond those previous reports by identifying the context trees characterizing the learned temporal dependencies.
	
Statistical learning paradigms have mostly relied on the detection of
the Mismatch Negativity, consisting in an electrophysiological
response induced by a deviant stimulus within a sequence of otherwise
regular stimuli
\cite{dehaene_neural_2015,winkler_evidence_2012,friston_theory_2005}. Likewise,
in our experiment the retrieving of some subtrees could be explained
by the presence of a Mismatch Negativity response. For instance, in
the Quaternary condition, consider the subtree $\{000,100,200\}$. The
silent unit 0 appearing as third element in each of these strings
appears as a surprise in two of the strings and is expected to occur
in the third string. In effect, the silent unit 0 was unexpected after
either 00 or 10 and was expected to occur after the string
20. Therefore, the distinction between the EEG segments collected
while the silent unit 0 appears after 00 versus its occurrence after
20 is a typical situation in which the Mismatch Negativity approach
would be successful in distinguishing the statistical features of the
two EEG signals. The same occurs when we compare the EEG segments
collected while the silent unit 0 appears after 10 versus its
occurrence after 20. So these are situations in which the Mismatch Negativity effect contributes to the context identification task.

Using the context tree approach, however, we went beyond the
identification of a Mismatch Negativity response. This is the case of
the frequently retrieved subtree $\{01,21\}$ in the Quarternary
condition. In effect, the weak beat 1 appearing as second element in
the two strings is expected to occur with high probability in both
cases. It occurs with high probability after the context 2 and it
occurs also with high probability either after 200 or after 10. In
these cases, the identification of a difference in the EEG collected
while the weak beat unit is presented either after context 2 or after
contexts 200 and 10 can not be explained by the mismatch negativity
effect. The successful retrieving of a context in these cases attests
that the volunteer is learning the structure of the sequence of
stimuli beyond the surprise level. This strongly suggests that the
participants succeeded in identifying the context tree generating the
stimuli sequence. This context tree identification would allow
predicting the upcoming stimulus
\cite{mcclelland_why_1995,henke_model_2010,norman_modeling_2003}.

It is interesting to observe that in some situations the Mismatch Negativity effect can mask the differences between the laws of the EEG segments. For instance, in the Ternary condition, our model predicts that the subtree $\{00, 10, 20\}$ should appear in the retrieved context tree. It turns out that this only occurs in 3 out of 18 electrodes. This is surprising since to well predict the next symbol it is crucial to distinguish between 00 or 10 in one side, and 20 in the other side. After each of the first two strings a strong beat always appear, while after 20 a weak beat will occur with high probability and a silent unit will occur with small probability. How can we explain that we did not succeed in identifying this distinction in the EEG data more often? This could be due to the fact that in this case the silent unit represented by the symbol 0 appearing at the end of the pair is always the consequence of erasing the most probable weak beat represented by the symbol 1. This is a typical situation of mismatch negativity in the EEG signal triggered by the omission response \cite{Wacongne2011}. Maybe this explains the difficulty in distinguishing the probability distributions generating the EEG segments associated to the leaves of this subtree.
	
Regarding the spatial distribution of the retrieved mode context trees in the scalp, we found that those corresponding exactly to the tree that generated the sequence of stimuli were mostly obtained in the frontal region. Correctly identifying all the contexts at a particular scalp position could indicate that the temporal dependencies characterizing the sequence of stimuli are preferably
encoded by the underlying brain networks. This is consistent with fMRI studies showing that the prefrontal cortex is involved in learning context-based statistics \cite{donoso_foundations_2014, Wang8412}. 

The results also show some variability regarding the context trees estimated across the scalp. One could suppose that several putative context tree models are being generated across the scalp and used to make predictions. Keeping multiple models is advantageous in volatile environments, where transition probabilities or even temporal dependencies change across time \cite{donoso_foundations_2014,konovalov_neurocomputational_nodate}.

A common issue in the analysis of functional data such as EEG segments
is the need of some type of normalization to compare quantities like
the maximum amplitude or the variance of the signals collected from
different participants. The context tree approach intrinsically
normalizes the data per participant and electrode. This is done by
extracting the essential features concerning equality of the
probability distributions. These features are then represented by the
retrieved context tree. Data from different participants are only put
together at the end through the mode context trees. Summarizing the
context trees retrieved per electrode and participant using the mode
context tree is a novelty introduced in the present paper.

The goal of the present article was to find experimental evidence that
it is possible to retrieve from EEG data the features of the
probabilistic algorithm governing the successive stimuli presented to a volunteer.  One could suppose that a large number of volunteers would be required to achieve this goal. Our results on the interindividual variability suggest however a quite homogeneous performance across participants and conditions.
 
A main concern refers to the minimum size of the sequence of stimuli required to consistently retrieve the context tree per participant, electrode and condition. First of all, it has been rigorously proven \cite{duarte_retrieving_2019} that our statistical procedure is consistent. A simulation study \cite{duarte_retrieving_2019} indicated that for sequences of auditory stimuli longer than 200 units, the proportion of correctly identified context trees exceeded 0.95. In our experimental protocol, the sequences of auditory stimuli have almost 800 units, which should be sufficiently large to well identify the context trees from the EEG data. Obviously, a major difference between the simulation study and the EEG data analysis is the much smaller variability of the diffusion processes used in the simulation study as proxies of the EEG segments \cite{duarte_retrieving_2019}. The intrinsic variability of the EEG data implies in a higher variance of the results and pushes towards using larger sequences of auditory stimuli. Increasing the length of the sequences of stimuli has however the drawback of longer experimental sessions \cite{Luck05}. The homogeneity of the results obtained across participants and contexts suggests that we achieved a good balance between the push towards longer sequences of stimuli and the limitations imposed by increasing the length of the experimental session. 
	
Humans are great at learning structured sequences of stimuli such as
those found in language and music \cite{dehaene_neural_2015}. Our
approach allowed retrieving from the collected
EEG signals the higher-order temporal dependencies present both in the
Ternary and the Quaternary conditions. Our results indicate that the
statistical regularities learned from a relatively short exposure to
these structured sequences of auditory stimuli become fully encrypted
in brain signals. Interestingly, the context trees governing the
generation of the sequences of the stimuli were retrieved mostly from
EEG signals acquired in frontal electrodes, suggesting that these
regions are more likely responsible for encoding higher-order temporal
dependencies characterizing the context trees driving the sequences of auditory stimuli.
	
\section*{Data Availability}
    
The EEG data and the source code used in the statistical analyses can be downloaded from \\  \url{https://neuromat.numec.prp.usp.br/neuromatdb/EEGretrieving/}.

\bibliography{brain_sensitivity_eeg_references}
	
\section*{Acknowledgements (not compulsory)}
	
This work is part of University of S\~ao Paulo project Mathematics, computation, language and the brain, FAPESP project Research, Innovation and Dissemination Center for Neuromathematics (grant 2013/07699-0), project Plasticity in the brain after a brachial plexus lesion (FAPERJ grants E26/010002902/2014, E26/010002474/2016) and Financiadora de Estudos e projetos FINEP (PROINFRA HOSPITALAR grant 18.569-8). Authors A.G and C.V. are partially supported by CNPq fellowships (grants 311 719/2016-3 and 309560/2017-9, respectively). Author C.D.V. is also partially supported by a FAPERJ fellowship (CNE 202.785/2018). Author A.D. was fully supported by CNPq and FAPESP fellowships (grants 201696/2015-0 and 2016/17791-9). Author G.O. was fully supported by CNPq and FAPESP fellowships (grants 201572/2015-0 and 2016/17789-4). Author N.H. was fully supported by a FAPESP fellowship (grant 2016/22053-7).
        
We thank Jorge Stolfi for many interesting discussions and constant help with computational issues. We also thank Raymundo Machado de Azevedo Neto for interesting discussions and help in the redaction of a preliminary manuscript version. 

\section*{Author contributions statement}
	
A.G. and C.D.V conceived the theoretical framework and the experimental protocol. A.D, G.O. and C.D.V. collected the data. A.D., R.F., A.G, N.H, and G.O, developed the statistical method used in this study. A.D. and N.H. analyzed the data. A.G., N.H., and C.D.V. wrote the article. All authors reviewed the manuscript.
	
\section*{Additional information}
	
\textbf{Competing interests}: The authors declare no conflict of interest.

\end{document}